\documentclass[preprint2]{aastex}

\shorttitle{Probing Radio Jet Collimation Regions}
\shortauthors{Ly, Walker, \& Wrobel}
\begin{document}

\title{An Attempt to Probe the Radio Jet Collimation Regions in NGC\,4278, NGC\,4374 (M84),
       and NGC\,6166}
\author{C. Ly,\altaffilmark{1} R. C. Walker, and J. M. Wrobel}
\affil{National Radio Astronomy Observatory,\altaffilmark{2} P.O. Box 0, Socorro, NM 87801}
\email{cly@u.arizona.edu,  cwalker@nrao.edu, jwrobel@nrao.edu}

\altaffiltext{1}{Present address: Steward Observatory, University of Arizona,
                 933 North Cherry Avenue, Tucson, AZ 85721-0065}
\altaffiltext{2}{The National Radio Astronomy Observatory is a facility of the
                 National Science Foundation, operated under cooperative
                 agreement by Associated Universities, Inc.} 
\begin{abstract}
NRAO Very Long Baseline Array (VLBA) observations of NGC\,4278, NGC\,4374 (M84),
NGC\,6166, and M87 (NGC\,4486) have been made at 43~GHz in an effort to image
the jet collimation region. This is the first attempt to image the first three
sources at 43~GHz using Very Long Baseline Interferometry (VLBI) techniques.
These three sources were chosen because their estimated black hole mass and
distance implied a Schwarzschild radius with large angular size, giving hope
that the jet collimation regions could be studied. Phase referencing was utilized
for the three sources because of their expected low flux densities. M87 was chosen
as the calibrator for NGC\,4374 because it satisfied the phase referencing
requirements: nearby to the source and sufficiently strong. Having observed M87
for a long integration time, we have detected its sub-parsec jet, allowing us to
confirm previous high resolution observations made by Junor, Biretta \& Livio,
who have indicated that a wide opening angle was seen near the base of the jet.
Phase referencing successfully improved our image sensitivity,
yielding detections and providing accurate positions for NGC\,4278, NGC\,4374
and NGC\,6166. These sources are point dominated, but show suggestions of extended
structure in the direction of the large-scale jets. However, higher sensitivity will
be required to study their sub-parsec jet structure.
\end{abstract}


\section{INTRODUCTION}
The high resolution capabilities of Very Long Baseline Interferometry
(VLBI) have given astronomers the ability to study jets in the inner
parsec of active galactic nuclei (AGN), leading to the possibility of
testing jet collimation theories and magnetohydrodynamic (MHD) simulations.
MHD simulations have been able to explain the formation of AGN jets by
magnetic fields threading through an accretion disk around a super-massive
black hole. The fields thread out from the disk in a rotating helical coil,
accelerating and guiding the plasma away from the disk. The resulting jet
structure is expected to show a large opening angle near the base and tight
collimation on larger scales \citep{meier01}. An example where the large
opening angle at the base of the jet may have been observed is the elliptical
galaxy, M87 \citep{junor99}. They reported that the jet must form within 30
Schwarzschild radii ($R_{\rm s}$) of the black hole and the collimation region
extends to beyond 100~$R_{\rm s}$ \citep{biretta02}. M87's close proximity,
large black hole mass, and bright jet make it arguably the best candidate to
study the jet collimation region.

Motivated by M87, we began an effort to observe other sources in which
the jet collimation region might be probed.  We selected sources based on
the expected angular size of the Schwarzschild radius
[$R_{\rm s}=(2GM_{\bullet})/c^2$], since that is likely to set the scale
of the collimation region. Hence, nearby galaxies with large black hole mass
were favored. We further selected sources based on the presence of detectable
radio emission.  Using black hole masses available in mid 2000, the most
promising sources were NGC\,4278, NGC\,4374 and NGC\,6166. The estimated black
hole masses available then were 15, 18 and 290$\times10^8~M_{\sun}$, respectively
\citep{geb00,mag98}. NGC\,6166 is relatively distant, but what turned out
to be an excessively large black hole mass estimate qualified it as part of our
sample.

A recent study finds that the correlation between the mass of the central black
hole in a galaxy and the velocity dispersion ($M_{\bullet}-\sigma$ relation) is
\begin{equation}
\log{(M_{\bullet}/M_{\sun})}=\alpha + \beta\log{(\sigma/\sigma_{\circ})}
\label{eqn1}
\end{equation}
where $\alpha$~=~8.13~$\pm$~0.06, $\beta$~=~4.02~$\pm$~0.32, $M_{\bullet}$
is the black hole mass, $\sigma$ is the central velocity dispersion in units of
km s$^{-1}$ and $\sigma_{\circ}$ is a reference velocity taken to be 200~km s$^{-1}$
\citep{tremaine02}. For our three target sources, the black hole masses computed from
Equation~\ref{eqn1} are substantially lower than those used for the selection process.

Current distance and black hole mass estimates are provided in Table~\ref{table1}
for our target sources and M87. The first and second columns list the NGC and
3C catalog numbers of the sources; Messier catalog names are in parentheses.
Column 3 provides the computed distances $D$ in units of Mpc to these sources
using surface brightness fluctuation (SBF) distance moduli, and Column 4 gives
the references for these values. Black hole masses are tabulated (Column 5) in
units of 10$^8$~$M_{\sun}$. These numbers were computed using Equation~\ref{eqn1}
and the velocity dispersions from the references (Column 6). For M87, the black
hole mass is as reported by \cite{tremaine02}. Using the values in Columns 3 and 5,
the inferred angular sizes $\theta_{\rm s}$ in micro-arcseconds ($\mu$as) of the
Schwarzschild radius is provided in Column 7.

Section 2 will describe our observing techniques, the calibration process and our
imaging approach. Section 3 will briefly introduce characteristics of each source,
including previous related observations, and discuss the implications inferred
from our new VLBA images. Finally, Section 4 will summarize our results, and remark
on the future prospects for probing the jet collimation regions.

\section{OBSERVATIONS, CALIBRATION, AND IMAGING}
Observations with the 10 VLBA \citep{napier94} antennas of NGC\,4374 (also M87), NGC\,6166
and NGC\,4278 were made on 2001 October 12, 2002 February 14, and 2002 September 19,
respectively. These observations were made at 43.127~GHz using 32~MHz bandwidth
in each polarization and 2 bit sampling. The resolution was typically 0.2 by
0.4~milli-arcseconds (mas). Since the target sources were faint, phase-referencing
techniques were used to improve image sensitivity. Suitable phase calibrators
(peak flux density greater than 100~mJy~beam$^{-1}$ and within 2$\arcdeg$~from target
sources) were J1221+2813 (W Comae) for NGC\,4278, M87 for NGC\,4374, and
J1635+3808 for NGC\,6166. For the first observation, NGC\,4374 and M87 were
each observed with short, scheduled scans of 15~seconds, using the ``nodding''
technique. Analysis of that observation showed that only a few seconds of on-source
time were obtained for each scan, and the coherence time was adequate to allow
a longer switching time. Therefore, the other two observations were made with scans
of 30~seconds on target and 20~seconds on calibrator, for a total switching time of
50~seconds. The approximate total integration times were 200~minutes NGC\,4278,
91~minutes for NGC\,4374, 85~minutes for M87, and 141~minutes for NGC\,6166.

In addition to our target and calibration sources, we also observed fringe finders
and check sources. The fringe finders, 3C\,345 and 3C\,279, were used to make delay
corrections. Check sources were included to examine the quality of the phase
referencing. The check sources were J1218+1105, J1217+3007 and 3C\,345 while
observing NGC\,4278, NGC\,4374 and NGC\,6166, respectively.

We followed the standard VLBA data reduction procedure using the NRAO Astronomical
Image Processing System (AIPS) as described in Appendix C of the AIPS Cookbook.
Delays were corrected using fringe fitting with the fringe finders. They were
sufficiently constant that no further fringe fitting was done. Phase calibrators
were iteratively imaged and self-calibrated. The images were then used as models
to determine the amplitude and phase gains for the antennas. These gains were
applied to the target and check sources (phase referencing) and images were made.
The target sources proved to be strong enough to allow slowly varying gain errors
to be corrected and improved images to be made using self-calibration with long
(10, 30 or 60~minutes) solution intervals. Saint Croix was excluded for the phase
referenced sources because of rapid phase fluctuations, but was used for the
self-calibrated images of the sources observed on 2001 October 12.

Observational results on all sources, excluding fringe finders, are provided in
Table~\ref{table2}. We have included both phase-referenced and self-calibrated
flux densities to indicate the quality of the phase referencing.
Columns 1 and 2 provide the name of the source and what it was used for during
the observation. `Calib', `Target' or `Check' indicates that the source was used
either as a calibrator, a target or a check source. Columns 3 and 4 list the
right ascension and its one-sigma error with the declination and its one-sigma
error below it. The coordinates and errors for the calibrators were obtained from
the International Celestial Reference Frame Extension 1
\citep[ICRF-Ext.1;][]{iers99,ma98}; the positions for the other sources were
measured from our images. Errors account for the error of the calibrator
position (the dominate source of error), an assumed 0.1~mas potential error due
to differential geometry and atmospheric model errors between calibrator and
target or check source, and errors in the measurement of position from the images.
Column 5 reports the angular offset ($\theta$) in degrees from the calibrator.
Columns 8 through 13 are the peak flux density per beam ($S_{\rm p}$) of the source
in mJy~beam$^{-1}$, the integrated flux density ($S_{\rm i}$) in mJy, and the
off-source image noise (RMS) in mJy~beam$^{-1}$ for both the phase-referenced and
the self-calibrated images. To obtain the integrated flux for our target sources
and M87, we used a region that is 2.5~mas~$\times$~1~mas for NGC\,4278,
2.8~mas~$\times$~1~mas for NGC\,4374, 2.5~mas~$\times$~5~mas for M87, and
1.5~mas~$\times$~1.5~mas for NGC\,6166. Phase-referenced and self-calibrated images
have been convolved to the same beam size and position angle listed in Column 5 and 6.
For 3C\,345, observed on 2002 February 14, only the portions of the data scheduled as a
check source were included in the phase referencing. The self-calibrated values include
all data. The uncertainty of the self-calibrated flux densities is difficult to
determine, but we estimate it to be about 10\%. It depends on the quality of the VLBA
a priori gain calibration, the ability of the phase referencing to remove short term
fluctuations for weak sources, the quality of the self-calibration, and the image noise.
In all cases, the flux scale (gain normalization) was set by observations above
30\arcdeg~elevation at a subset of antennas that were well-behaved and had good
weather on the day of the observation.

The 100-$\mu$as potential error in the positions due to differential model
errors is approximate and was derived in two ways. First, the raw phases showed
approximately 10 turns of phase across the day, which is equivalent to across the
sky on the longest baselines. Scaling this to an angular error, and decreasing it
by the ratio of the calibrator/source separation to a radian, gives an estimate
of about 0.1~mas. Second, \cite{pradel03} have done a study of the errors in
differential VLBI astrometry, suggesting errors of approximately 0.1~mas for our
observations (Lestrade, private communication). Two of our check sources
(J1218+1105, 3C\,345) and two of our target sources (NGC\,4278, NGC\,4374) are
included in the ICRF-Ext.1. In all cases, the calibrator-target or calibrator-check
source differential position measured from our images agrees with the differential
positions from the ICRF-Ext.1 to better than one-third to one-half of the reported
ICRF-Ext.1 errors. This confirms that our
measured positions are reasonable and implies that both our differential positions,
and the relative positions of close pairs of sources in the ICRF-Ext.1, are
significantly more accurate than the absolute position errors in the ICRF-Ext.1.

\section{DISCUSSION}

\subsection{NGC\,4278, NGC\,4374 and NGC\,6166}
\begin{figure}[htp]
\vspace{18cm}
\includegraphics{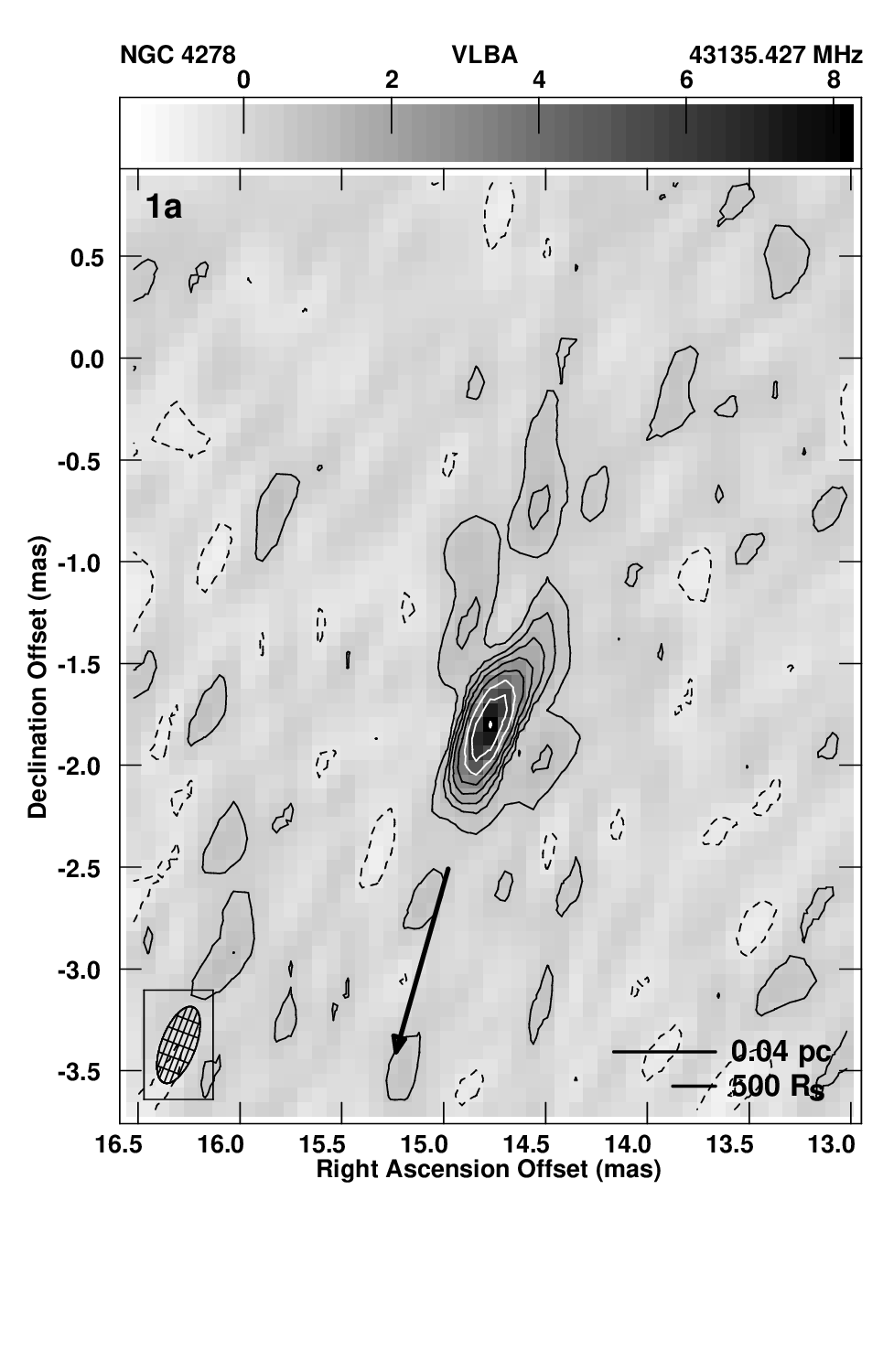}
\includegraphics{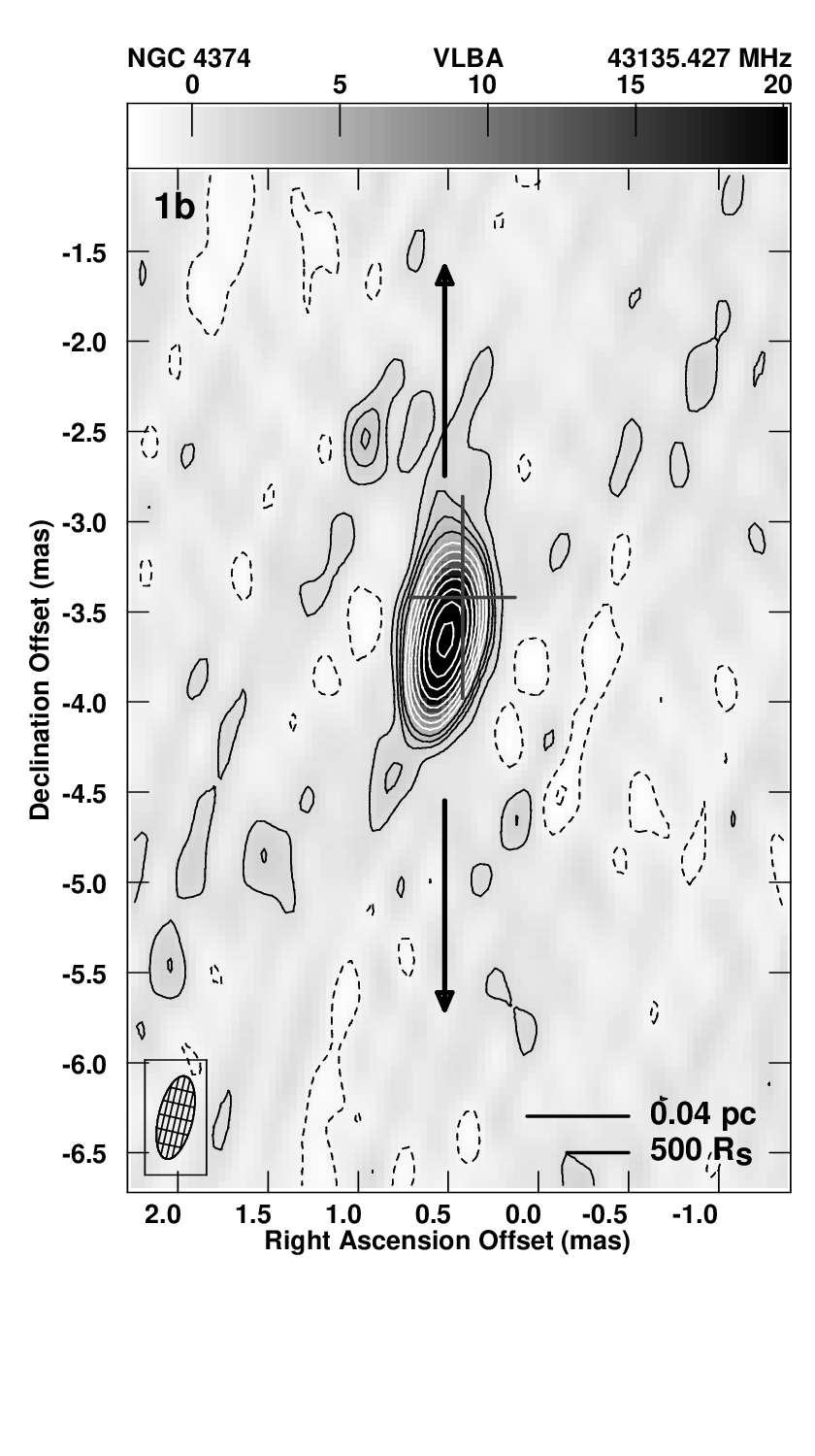}
\includegraphics{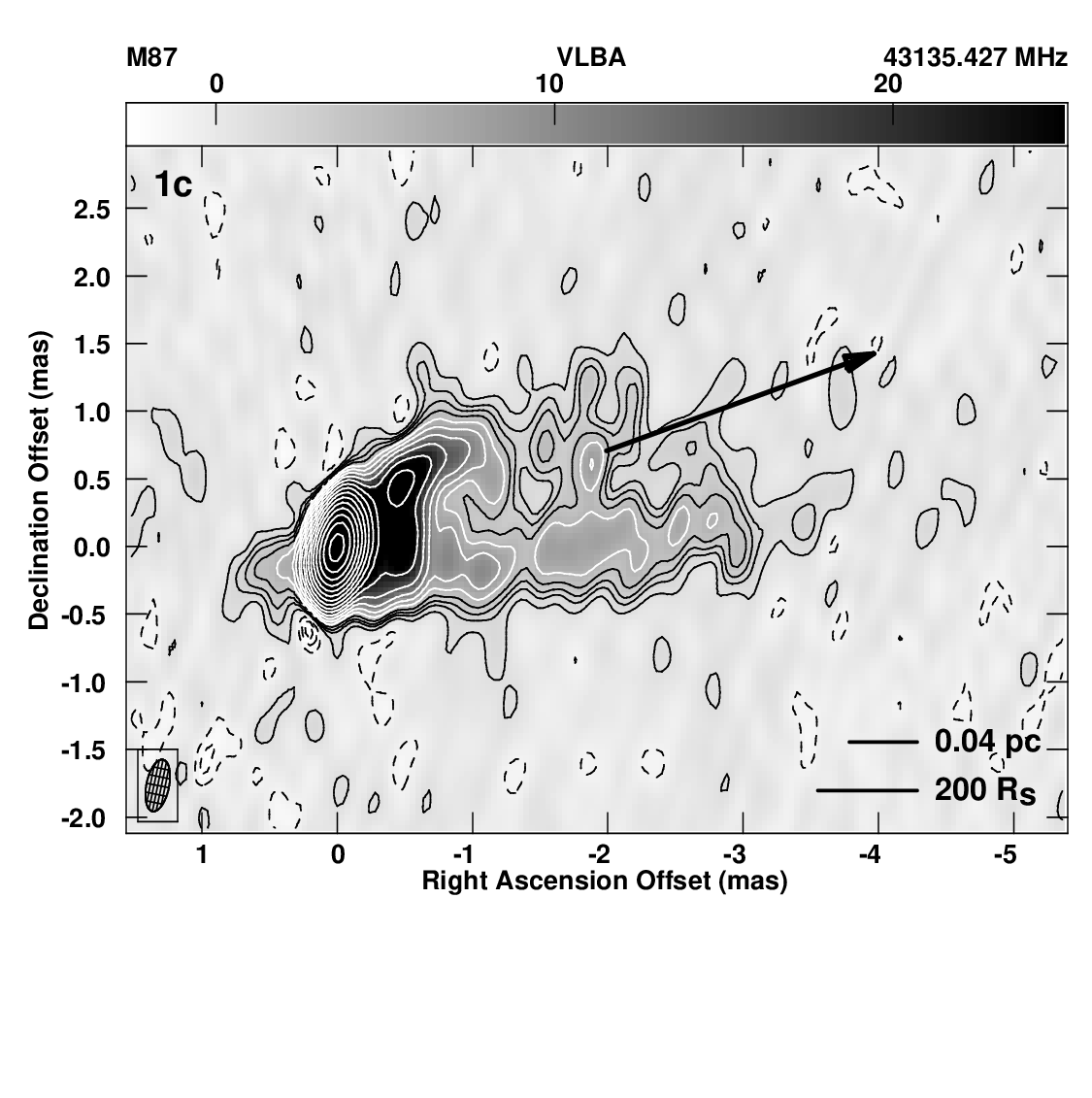}
\includegraphics{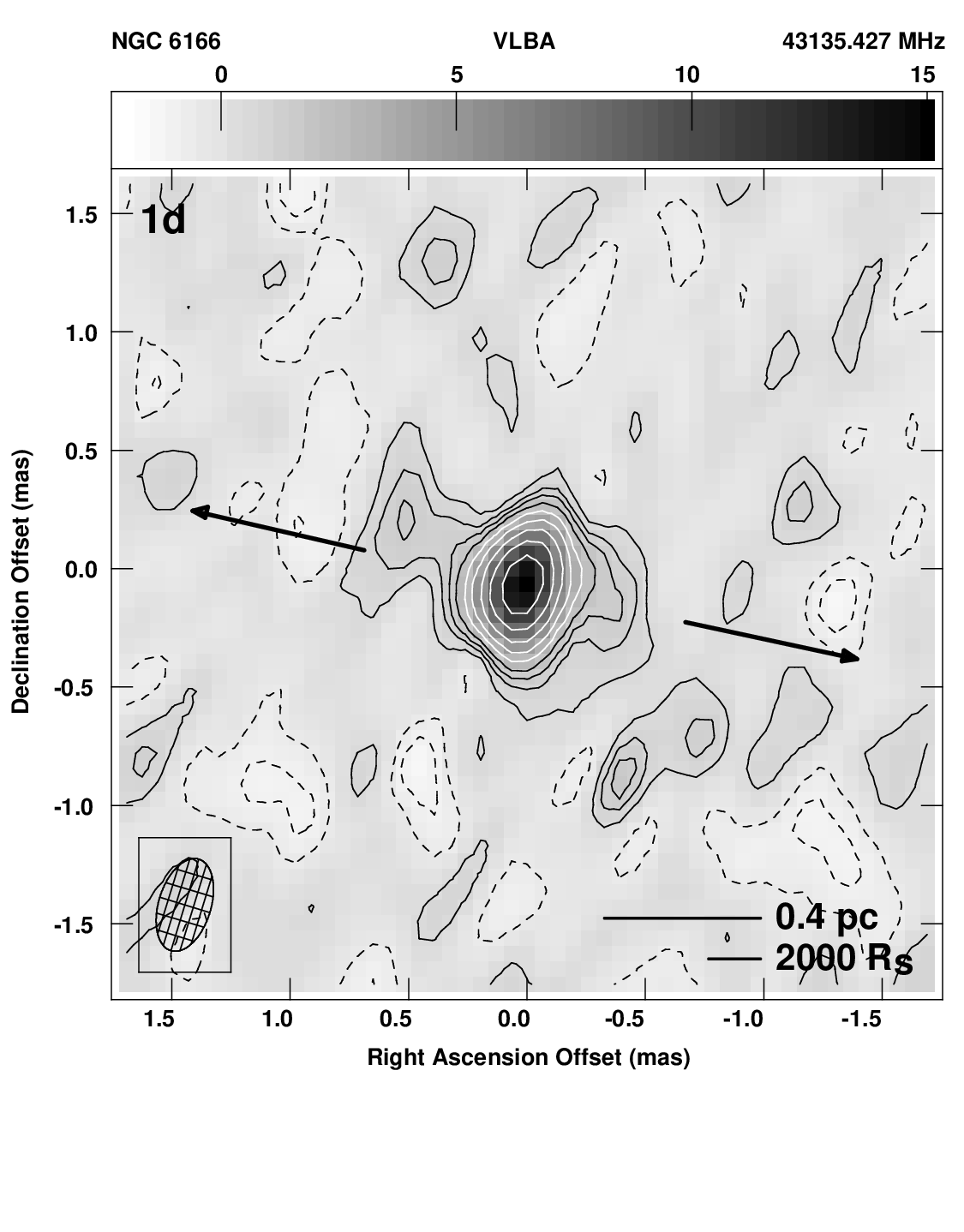}
{{}}
\caption{VLBA images of Stokes I at 43~GHz. Contour levels are scaled from the lowest
level by -2.8, -2, -1, 1, 2, 2.8, 4, 5.7, 8 and multiples of $\sqrt{2}$ thereafter.
({\em 1a}) NGC\,4278: lowest level is .5~mJy~beam$^{-1}$; grayscale levels range from -1.4 to 8.2~mJy~beam$^{-1}$.
({\em 1b}) NGC\,4374: lowest level is 1~mJy~beam$^{-1}$; grayscale levels range from -1.9 to 20~mJy~beam$^{-1}$.
({\em 1c}) M87: lowest level is 1~mJy~beam$^{-1}$; grayscale levels range from -2.5 to 25~mJy~beam$^{-1}$.
({\em 1d}) NGC\,6166: lowest level is 0.5~mJy~beam$^{-1}$; grayscale levels range from -2.0 to 15~mJy~beam$^{-1}$.
The arrows indicate the direction of the jets seen in low frequency VLBI observations.
The position offsets are relative to the a priori positions used in correlation, except for
NGC\,6166 where position corrections were made. The cross in {\it 1b} marks the ICRF-Ext.1
position. The beam sizes and peak flux densities (self-calibrated) can be found in
Table~\ref{table2}. The apparent counterjet in M87 is still questionable since it may be an imaging artifact.}
\label{fig1}
\end{figure}

Our VLBA images of NGC\,4278 (Figure~\ref{fig1}a), NGC\,4374 (Figure~\ref{fig1}b),
and NGC\,6166 (Figure~\ref{fig1}d) are currently the highest resolution images
available for these galaxies. Continuum emission has been detected from the cores
at 43~GHz with FWHM resolutions of about 400~$\mu$as $\times$ 200~$\mu$as. While
these VLBA images do not clearly reveal any jet structures, they do provide
indications of low-level emission that may be associated with jets traced at lower
frequencies and lower resolutions. Before discussing each galaxy below, we note
that the photometry of the cores at sub-mas resolution (Table~\ref{table2}) may
help constrain spectral energy distributions of any advection-dominated accretion
flows in these galaxies \citep{matt101,matt201,kleijn02}.

\subsubsection{NGC\,4278}
Continuum studies of this elliptical galaxy at arcsecond resolution show dominance
by a flat-spectrum, time-variable component at frequencies of 1-15~GHz
\citep{wrobel84,wrobel91,nagar01,nagar02}. At a FWHM resolution of 3$\arcsec$~(230~pc),
this component is linearly polarized at 8.4~GHz by 0.34~$\pm$~0.06 percent with an
electric vector PA of 16~$\pm$~7$\arcdeg$~\citep{bower02}. Assuming Faraday rotation
effects are minimal at this frequency, this polarized emission probably arises in the
diffuse structure, mainly to the south and southeast, traced in images at 8.4~GHz at
a FWHM resolution of 200~mas \cite[16~pc;][]{wilk98} and at 5~GHz at FWHM resolutions
of 400~mas \citep[31~pc;][]{wrobel84} and 2.5~mas \citep[0.2~pc;][]{falcke00,gio01}.
In Figure~\ref{fig1}a, the arrow indicates the direction to the jet component offset
from the 5~GHz core by 10~mas (0.8~pc), as measured with the VLBA by \citet{falcke00}.
While our VLBA image of NGC\,4278 at 43~GHz shows hints of low-level emission mainly
to the north, that emission is too close to the noise to be considered reliable.
Finally, we note that our VLBA image at 43~GHz recovers less than one half of the
43-GHz flux density of the unresolved VLA core, in 2000 February, at a FWHM resolution
of about 500~mas \citep[39~pc;][]{matt101}. Such behavior is expected because the VLBA
is insensitive to mas-scale emission at 43~GHz, known to be present at lower
frequencies.  Still, some of this difference could originate through variability of
the sub-mas core between 2000 February and 2002 September.

\subsubsection{NGC\,4374 (M84)}
This elliptical galaxy hosts a well-studied Fanaroff-Riley class I (FR-I) radio continuum
source \citep{laing87}. Images at 5~GHz at a FWHM resolution of 500~mas (44~pc) show
two large-scale jets that are initially asymmetric, with the northern main jet being
brighter than the southern counterjet. Images at 1.7~GHz at FWHM resolutions of 11~mas
\citep[1~pc;][]{xu00} and of 6~mas \citep[0.5~pc;][]{gio01} show extensions to the north.
A VLBA image at 5~GHz at a FWHM resolution of 2.7~mas (0.2~pc) shows a core plus two
components to the north and one to the south, offset in the directions indicated by
the arrows in Figure~\ref{fig1}b \citep{nagar02}. The available ICRF-Ext.1 position and
1-$\sigma$ error for NGC\,4374 are marked by a cross in Figure~\ref{fig1}b
\citep{iers99}. Our VLBA image of NGC\,4374 at 43~GHz shows that the core appears to be
slightly extended to the north. Unfortunately, the image sensitivity for this galaxy was
reduced compared to the other two target sources, due to our less efficient observing
strategy. For this reason, and because it is strong enough not to require phase
referencing, NGC\,4374 is a prime choice for future observations with significantly
improved sensitivity.

\subsubsection{NGC\,6166 (3C\,338)}
This cD galaxy with multiple nuclei hosts an FR-I source
\citep[plus references therein]{gio98}. VLBI images between 1.7 and 8.4~GHz with FWHM
resolutions as high as 1.2~mas (0.7~pc) reveal two-sided parsec structure in the
direction of the arrows in Figure~\ref{fig1}d. Our VLBA image of NGC\,6166 at 43~GHz
shows slight elongations along those directions. It is unfortunate that the black-hole
mass estimate has been revised downward by a factor of 40, offering no real prospect for
probing the jet collimation region with this VLBA observation.

\subsection{M87}

This elliptical galaxy hosts an FR-I source whose main jet is
especially well-studied on arcsecond and milli-arcsecond scales
\citep[plus references therein]{biretta02}.  Global VLBI observations
at 43~GHz offer sub-mas resolution and indicate that the initial jet
collimation region extends to at least 0.5~mas \citep{junor99},
corresponding to 0.04~pc or 140~$R_{\rm s}$ for the black hole mass
and galaxy distance in Table~\ref{table1}.  Figure~\ref{fig1}c shows
our VLBA image of M87 at 43~GHz, with a FWHM resolution of 390~$\mu$as
$\times$ 170~$\mu$as, which independently confirms the Junor et al.
result of collimation over the inner 1~mas. Moreover, jet emission is
detected out to about 3~mas (0.2~pc or 820~$R_{\rm s}$) along the
direction of the large-scale jet indicated by the arrow in
Figure~\ref{fig1}c. One difference between our image and that of Junor
et al. is the presence of a weak counterjet in our image.  Detection
of a counterjet would be very interesting, if real.  But such
structures are also a natural consequence of amplitude calibration
errors.  We attempted, through manipulation of the region in which the
CLEAN algorithm was allowed to place components during the
self-calibration/imaging iterations, to make images without the
counterjet.  While successful, the counterjet tended to reappear when
the CLEAN boxes were opened up to allow it.  In the end, despite
considerable testing, the counterjet detection can only be considered
suggestive, not reliable.

\subsection{Other sources}
One of our calibrators, J1221+2813 (W Comae), showed extended structure with a
component 3~mas southeast from the bright core. The structure is consistent with that
seen at 22~GHz by \cite{wiik01}. Our observation of 3C\,345 shows a knot 1~mas west
from the radio core with additional extended emission. This is consistent with
22~GHz observations made by \cite{ros00}. J1218+1105 appears to be a compact source at
43~GHz. The check source for our third observation (J1217+3007) is seen mostly as a
compact source with low-level emission to the west. Our brightest calibrator, J1635+3808
has a minor jet to the west, appearing ``thumb-like''.

\section{CONCLUSION}
VLBI observations at 43~GHz have shown that M87 jet has a wide opening angle
in the inner $100~R_{\rm s}$, providing important clues to jet collimation mechanisms
\citep{biretta02}. The VLBA observations reported here were the first
from an effort to identify other sources in which the structure of the base of
the jet might be observed. The target sources were chosen to have radio continuum
emission and large, angular-size black holes. Each of the three target sources,
NGC\,4278, NGC\,4374, and NGC\,6166 (3C\,338), had point-dominated structure
with only hints of the beginning of the jet. The sensitivity of the observations
proved to be inadequate to determine the structure of the jet bases. M87 was also
observed as the calibrator for NGC\,4374; the resulting image agrees with the
results of \cite{junor99}. Future observations will concentrate on improving
sensitivity and finding new sources.

The observations reported here involved phase referencing at 43~GHz with the
VLBA. The technique worked well with switching angles between calibrator and
target of 1-2$\arcdeg$~and switching time intervals of 50~seconds (20~seconds on
calibrator, 30~seconds on target). A 30~second switching time (i.e. 15~seconds
per scan) was tried, but gave unreasonably short on-source integration and
was not required by the coherence time. The methods used here allow detection
of sources as weak as 2~mJy at 43~GHz with $\sim8$-hour observations.

\acknowledgments
This research has made use of the NASA/IPAC Extragalactic Database, which is
operated by the Jet Propulsion Laboratory, Caltech, under contract with the
National Aeronautics and Space Administration, and NASA Astrophysics Data
System. C. Ly acknowledges support at the National Radio Astronomy Observatory
by the Research Experiences for Undergraduates program of the National Science
Foundation.

\clearpage
\newpage
\begin{deluxetable}{llccccc}
\tablenum{1}
\tablecolumns{7}
\tablewidth{0pc} 
\tablecaption{Distance and Black Hole Mass Estimates}
\tablehead{
\colhead{(1)}&\colhead{(2)}&\colhead{(3)}  &\colhead{(4)} &\colhead{(5)}             &\colhead{(6)} &\colhead{(7)}       \\
\colhead{}   &\colhead{}   &\colhead{$D$}  &\colhead{}    &\colhead{$M_{\bullet}$}   &\colhead{}    &\colhead{$\theta_{\rm s}$}\\
\colhead{NGC}&\colhead{3C} &\colhead{(Mpc)}&\colhead{Ref.}&\colhead{(10$^8M_{\sun}$)}&\colhead{Ref.}&\colhead{($\mu$as)} }
\startdata
4278 \nodata    &\nodata& 16.1 & 1 & 3.4$^{+.6}_{-.5}$  & 2 & 0.42 \vspace{.15cm}\\
4374(M84)\nodata& 272.1 & 18.4 & 1 & 6.3$^{+1.3}_{-1.1}$& 2 & 0.68 \vspace{.15cm}\\
4486(M87)\nodata& 274   & 16.1 & 1 & 30$^{+10}_{-10}$   & 3 & 3.67 \vspace{.15cm}\\
6166 \nodata    & 338   & 125  & 4 & 7.1$^{+1.5}_{-1.2}$& 2 & 0.11 \vspace{.15cm}\\
\enddata
\label{table1}
\tablerefs{(1) Tonry et al. 2001; (2) Ravindranath, Ho, \& Filippenko 2002; (3) Tremaine et al. 2002; (4): Jensen et al. 2001}
\end{deluxetable}

\clearpage
\newpage
\begin{deluxetable}{llrlccc r@{}l r@{}l r@{}l c r@{}l r@{}l r@{}l}
\tablenum{2}
\tablecolumns{20}
\tablewidth{0pc}
\tablecaption{VLBA Astrometry and Photometry at 43\,GHz}
\tabletypesize{\small}
\rotate
\tablehead{
\colhead{(1)}   &  \colhead{(2)}  & \colhead{(3)}           & \colhead{(4)}                 & \colhead{(5)}        &\colhead{(6)}               &\colhead{(7)}        &
\multicolumn{2}{c}{(8)}&\multicolumn{2}{c}{(9)}&\multicolumn{2}{c}{(10)}               &\colhead{}&\multicolumn{2}{c}{(11)}&\multicolumn{2}{c}{(12)}&\multicolumn{2}{c}{(13)}\\
\colhead{}      & \colhead{}      & \colhead{RA ($\alpha$)} & \colhead{}                    & \colhead{}           &\colhead{}                  &\colhead{}           &
\multicolumn{6}{c}{Phase-referenced}                                                   &\colhead{}&\multicolumn{6}{c}{Self-calibrated}\\
\cline{8-13} \cline{15-20}
\colhead{Source}& \colhead{Source}& \colhead{DEC ($\delta$)}& \colhead{$\Delta\alpha$ (ms)} & \colhead{$\theta$}   &\colhead{FWHM}              &\colhead{PA}         &
\multicolumn{2}{c}{$S_{\rm p}$}&\multicolumn{2}{c}{$S_{\rm i}$}&\multicolumn{2}{c}{RMS}&\colhead{}& \multicolumn{2}{c}{$S_{\rm p}$}&\multicolumn{2}{c}{$S_{\rm i}$}&\multicolumn{2}{c}{RMS}\\
\colhead{name}  & \colhead{usage} & \colhead{(J2000)}       & \colhead{$\Delta\delta$ (mas)}& \colhead{($\arcdeg$)}&\colhead{(mas~$\times$~mas)}&\colhead{($\arcdeg$)}&
\multicolumn{2}{c}{(mJy)}&\multicolumn{2}{c}{(mJy)}&\multicolumn{2}{c}{(mJy)}          &\colhead{}&\multicolumn{2}{c}{(mJy)}&\multicolumn{2}{c}{(mJy)}&\multicolumn{2}{c}{(mJy)} }
\startdata
\multicolumn{20}{l}{Observed on 2001 October 12}\\
NGC\,4486\nodata & Calib & 12 30 49.423381 &0.017& \nodata&.39$\times$.17&168&\multicolumn{2}{c}{\nodata}&\multicolumn{2}{c}{\nodata}&\multicolumn{2}{c}{\nodata}&& 610&  &1300&  & 0&.49\\
                 &       & +12 23 28.04393 &0.25 &        &              &   &                           &                           &                           &&    &  &    &  &  &   \\
NGC\,4374\nodata & Target& 12 25 03.743336 &0.018&   1.5  &.47$\times$.19&167&                      45&  &                      80&  &                      0&.86&&  71&  & 100&  & 0&.56\\
                 &       & +12 53 13.13934 &0.27 &        &              &   &                           &                           &                           &&    &  &    &  &  &   \\
J1218+1105\nodata& Check & 12 18 26.092289 &0.018&   3.3  &.66$\times$.23&167&                      41&  &                      69&  &                      3&.2 &&  69&  &  95&  & 1&.8 \\
                 &       & +11 05 05.26293 &0.27 &        &              &   &                           &                           &                           &&    &  &    &  &  &   \\\tableline
\multicolumn{20}{l}{Observed on 2002 February 14}\\
J1635+3808\nodata& Calib & 16 35 15.492973 &0.055& \nodata&.38$\times$.21&159&\multicolumn{2}{c}{\nodata}&\multicolumn{2}{c}{\nodata}&\multicolumn{2}{c}{\nodata}&&5100&  &5600&  & 1&.6 \\ 
                 &       & +38 08 04.50060 &0.66 &        &              &   &                           &                           &                           &&    &  &    &  &  &   \\
NGC\,6166 \nodata& Target& 16 28 38.244693 &0.056&   1.9  &.40$\times$.22&163&                      14&  &                      22&  &                      0&.43&&  15&  &  23&  & 0&.44\\ 
                 &       & +39 33 04.23414 &0.67 &        &              &   &                           &                           &                           &&    &  &    &  &  &    \\
3C\,345   \nodata& Check & 16 42 58.809985 &0.056&   2.2  &.41$\times$.21&156&                    1400&  &                    3900&  &                     39&   &&1500&  &4600&  & 2&.8 \\
                 &       & +39 48 36.99432 &0.67 &        &              &   &                           &                           &                           &&    &  &    &  &  &   \\\tableline
\multicolumn{20}{l}{Observed on 2002 September 19}\\
J1221+2813\nodata& Calib & 12 21 31.690515 &0.019& \nodata&.42$\times$.17&158&\multicolumn{2}{c}{\nodata}&\multicolumn{2}{c}{\nodata}&\multicolumn{2}{c}{\nodata}&&  88&  & 280&  & 0&.5 \\
                 &       & +28 13 58.50026 &0.29 &        &              &   &                           &                           &                           &&    &  &    &  &  &   \\
NGC\,4278 \nodata& Target& 12 20 06.825429 &0.021&   1.1  &.40$\times$.17&159&                       6&.2&                      19&  &                      0&.32&&   8&.2&  18&  & 0&.32\\
                 &       & +29 16 50.71418 &0.31 &        &              &   &                           &                           &                           &&    &  &    &  &  &   \\
J1217+3007\nodata& Check & 12 17 52.081977 &0.021&   2.0  &.42$\times$.17&159&                      67&  &                     130&  &                      3&.5 &&  89&  & 150&  & 2&.8 \\
                 &       & +30 07 00.63590 &0.31 &        &              &   &                           &                           &                           &&    &  &    &  &  &   \\
\enddata
\label{table2}
\end{deluxetable}

\clearpage
\newpage
\onecolumn
\end{document}